\documentclass[prl,twocolumn,showpacs,preprintnumbers,amsmath,amssymb]{revtex4}
\usepackage{graphicx}
\usepackage{dcolumn}
\usepackage{amssymb}
\usepackage{amsmath}

\begin{document}

\title{Band Offsets at Semiconductor-Oxide Interfaces from Hybrid Density Functional Calculations}

\author{Audrius Alkauskas}
\author{Peter Broqvist}
\author{Fabien Devynck}
\author{Alfredo Pasquarello}

\affiliation{ Institute of Theoretical Physics, Ecole Polytechnique
F\'ed\'erale de Lausanne (EPFL), CH-1015 Lausanne, Switzerland}

\affiliation{Institut Romand de Recherche Num\'{e}rique en Physique
des Mat\'{e}riaux (IRRMA), CH-1015 Lausanne, Switzerland}

\begin{abstract}
Band offsets at semiconductor-oxide interfaces are determined through a scheme based on 
hybrid density functionals, which incorporate a fraction $\alpha$ of Hartree-Fock exchange.
For each bulk component, the fraction $\alpha$ is tuned to reproduce the experimental 
band gap, and the conduction and valence band edges are then located with respect to 
a reference level. The lineup of the bulk reference levels is determined through an interface 
calculation, and shown to be almost independent of the fraction $\alpha$. 
Application of this scheme to the Si-SiO$_2$, SiC-SiO$_2$, and Si-HfO$_2$ interfaces 
yields excellent agreement with experiment.
\end{abstract}

\date{\today}


\maketitle


The discontinuity in the local band structure at semiconductor-semiconductor and semiconductor-oxide interfaces
is a crucial physical property for the operation of most electronic and optoelectronic devices \cite{Sze}.
Early theoretical research \cite{Tersoff_1987}, mainly on semiconductor heterojunctions, provided
a deep understanding of the processes that govern the band alignments at interfaces \cite{Frensley_PRB_1977,
Harrison_JVST_1977,Tejedor_JPC_1978,Tersoff_PRB_1984}. This resulted in the development of 
theoretical models providing a reasonable description of band offsets and being particularly useful when a 
wide class of materials needs to be screened \cite{Robertson_JVST_2000}. However, these models mainly rely 
on bulk properties of the interface components \cite{Tersoff_1987}, and therefore do not account for the detailed 
atomic and electronic properties at the interface, which are known to affect band offsets \cite{Peacock_PRL_2004}.

Density functional calculations of band offsets provide a qualitative improvement by describing the 
electronic and atomic arrangements at the interface in a self-consistent way
\cite{Baraff_PRL_1977,Pickett_PRB_1978,VanDeWalle_PRB_1987,Wei_APL_1998,Peressi_JPD_1998}.
However, the most common approximations to the exchange-correlation energy, i.e.\ the generalized-gradient 
approximation and the local density approximation, lead to significant underestimations of band gaps,
thereby impairing the reliability of calculated band offsets.
While valence band offsets at semiconductor heterojunctions are described with reasonable accuracy 
due to cancellation of errors in both interface components, band-offsets errors for 
semiconductor-oxide interfaces can reach several eV \cite{Peacock_PRL_2004,Giustino_PRL_2005,Devynck_PRB_2007,Godet_APL_2007}.
Calculations based on the $GW$ perturbation theory yield accurate band offsets 
at interfaces due to an improved description of bulk band gaps \cite{Zhang_SSC_1988,Shaltaf_PRL_2008}. 
However, these calculations are computationally demanding and can only be applied to relatively small 
systems. For instance, the study of realistic semiconductor-oxide interfaces in which the oxide 
is noncrystalline \cite{Giustino_PRL_2005,Devynck_PRB_2007,Broqvist_APL_2008} is severely hindered.

In this work, we introduce a scheme for calculating band offsets at interfaces through the use of hybrid 
density functionals. These functionals incorporate a fraction $\alpha$ of Hartree-Fock exchange \cite{Becke_JCP_1993} 
and substantially improve the description of bulk band gaps \cite{Muscat_CPL_2001}.
We apply our scheme to model structures of the Si-SiO$_2$, Si-HfO$_2$, and SiC-SiO$_2$ interfaces, which all 
feature a realistic description of the complex transition region. The band structures of the two interface 
components are lined up through their reference levels at the interface \cite{VanDeWalle_PRB_1987}.
For each component, we perform bulk calculations tuning the fraction $\alpha$ to reproduce the experimental 
band gap. The lineup of the reference potential in the interface model is found to only weakly depend 
on the fraction $\alpha$ conferring consistency on our scheme. For the three interfaces studied, the 
calculated band offsets are in excellent agreement with experiment.


We considered a class of hybrid density functionals based on the 
generalized gradient approximation of Perdew, Burke, and Ernzerhof (PBE) \cite{PBE}, which 
are obtained by replacing a fraction $\alpha$ of PBE exchange with Hartree-Fock exchange \cite{PBE0}.
The functional defined by $\alpha$=0.25 is referred to as PBE0 and is supported by
theoretical considerations \cite{PBE0}.
Core-valence interactions were described through normconserving
pseudopotentials generated at the PBE level. The valence wave functions were expanded in a 
plane-wave basis set defined by an energy cutoff of 70 Ry. 
The interface calculations corresponding to large supercells were performed with a 
Brillouin-zone sampling restricted to the $\Gamma$ point. In the bulk calculations, 
the positions of the band extrema were determined through converged $k$-point samplings.
The integrable divergence of the Hartree-Fock exchange term was explicitly treated \cite{Gygi_PRB_1986}. 
Structural relaxations were carried out at the PBE level. We used the implementations 
in \textsc{q}uantum-\textsc{espresso} \cite{QE} and \textsc{cpmd} \cite{CPMD}. 


\begin{table}[t]
\caption{Band gaps (in eV) of Si, SiC, HfO$_2$, and SiO$_2$, calculated using functionals incorporating 
a varying fraction $\alpha$ of Hartree-Fock exchange: $\alpha$=0 (PBE), $\alpha$=0.25 (PBE0), 
and an optimal fraction $\alpha_0$ reproducing the experimental band gap.}
\begin{ruledtabular}
\begin{tabular}{l c c c c}
          &  PBE   &  PBE0   & Optimal ($\alpha_0$) & Expt.  \\ 
\hline
Si        &  0.6   &  1.8    & 1.1 (0.11)          & 1.1  \\
SiC       &  2.2   &  3.9    & 3.3 (0.15)          & 3.3  \\ 
HfO$_2$   &  4.3   &  6.7    & 5.9 (0.15)          & 5.9  \\
SiO$_2$   &  5.4   &  7.9    & 8.9 (0.34)          & 8.9  \\
\end{tabular}
\label{tab1}
\end{ruledtabular}
\end{table}

The choice of the model structures requires particular attention. Indeed, previous density-functional 
studies on crystalline-crystalline Si-ZrO$_2$ and SiO$_2$-HfO$_2$ interfaces have revealed a strong 
sensitivity of the band offsets on the adopted model of the interfacial bonding pattern \cite{Peacock_PRL_2004}. 
Therefore, we considered model interfaces in which the oxide is amorphous to ensure that 
the transition region is smooth in terms of bond parameters and coordination 
\cite{Giustino_PRL_2005,Devynck_PRB_2007,Broqvist_APL_2008}. For the Si-SiO$_2$ interface, such a choice
led to variations of 0.1 to 0.2 eV in the band offsets calculated at the PBE level \cite{Giustino_APL_2005}.
The Si-SiO$_2$ interface adopted here is described through a 217-atom superlattice model in which layers of crystalline 
Si (9 monolayers) and of amorphous SiO$_2$ (17 \AA) alternate \cite{Giustino_PRL_2005}. This interface model structure 
incorporates a set of atomic-scale features inferred from experimental data. The SiC-SiO$_2$ interface is modeled by 
a 237-atom slab characterized by a chemically abrupt transition between crystalline 4$H$-SiC (8 monolayers) and amorphous SiO$_2$ 
(16 \AA) \cite{Devynck_PRB_2007}. The Si-HfO$_2$ interface is described by a superlattice model comprising 282 atoms 
and including a SiO$_2$ interlayer (7 \AA) between crystalline silicon (10 monolayers) and amorphous HfO$_2$ (12 \AA) \cite{Broqvist_APL_2008}.
To represent the bulk of the interface components, we used the corresponding crystalline structures for Si and 4$H$-SiC, 
a disordered model for SiO$_2$ \cite{Sarnthein_PRL_1995}, and the monoclinic structure for HfO$_2$ \cite{Broqvist_APL_2006}. 
In Table \ref{tab1}, the band gaps of these reference bulk components are calculated at the
PBE and PBE0 levels and compared to experimental values. As well known, the PBE band gaps severely underestimate
their measured counterparts. While the inclusion of Hartree-Fock exchange always enhances the calculated band gap,
the comparison with experiment is not systematically improved at the PBE0 level.


\begin{figure}
\includegraphics[width=7.5cm]{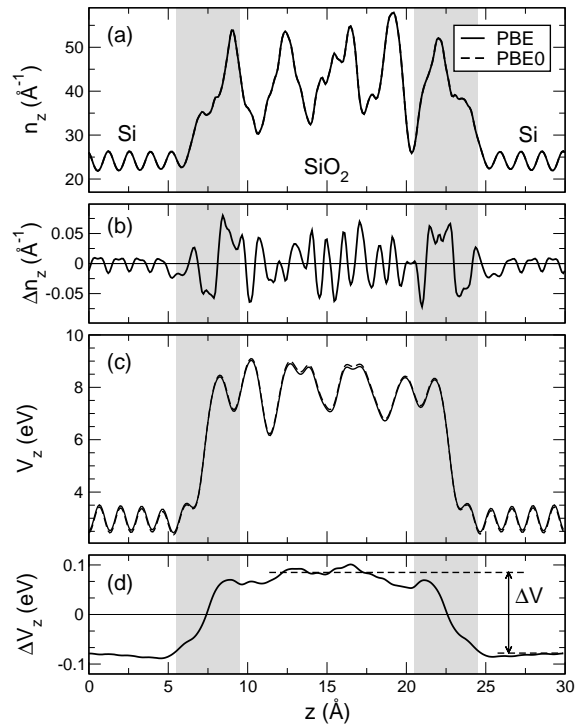}
\caption{
Planar-averaged (a) electron density  and (c) local 
potential across the Si-SiO$_2$ interface calculated in PBE (solid) and PBE0 (dashed).
The difference between the electron densities and the local potentials in the two schemes 
is shown in (b) and (d), respectively.  The shaded areas correspond to the transition 
regions between Si and SiO$_2$.}
\label{fig1}
\end{figure}

First, we calculated band offsets at the three interfaces within both PBE and PBE0.
The bulk band edges were aligned through a reference potential calculated across the interface \cite{VanDeWalle_PRB_1987}. 
As reference potential, we generally used the local potential \cite{note1}, but for SiO$_2$ and HfO$_2$ we
resorted to the energy levels of the deep O $2s$ and Hf $5s$ states which are less sensitive to structural disorder.
Focusing on the Si-SiO$_2$ interface, we show in Fig.\ \ref{fig1}  
the planar-averaged electron density and local potential across the interface.
With respect to PBE, PBE0 only yields a small redistribution of the electron density [Fig.\ \ref{fig1}(b)], 
which results in a difference of $\Delta V$=0.16 eV between the lineups of the potentials at the interface [Fig.\ \ref{fig1}(d)].
This indicates that the dipole contribution to the band offsets is already well described at the 
PBE level. Nevertheless, band offsets calculated at the PBE0 level noticeably improve upon the PBE ones. 
For instance the valence band offset goes from 2.5 eV to 3.3 eV, to be compared with the experimental 
value of 4.4 eV \cite{Himpsel_PRB_1988}.  This improvement is mostly due to a better description 
of bulk band gaps in PBE0 (Table \ref{tab1}). However, deviations with respect to experiment are still 
remarkable. Similar observations also hold for the other interfaces (Table \ref{tab2}).
 
\begin{table}
\caption{Valence ($\Delta E_{\text{v}}$) and conduction ($\Delta E_{\text{c}}$) band offsets at the Si-SiO$_2$, SiC-SiO$_2$, 
and Si-HfO$_2$ interfaces calculated in PBE, PBE0, and the mixed scheme. Experimental band offsets  
are from Refs.\ \cite{Himpsel_PRB_1988}, \cite{Afanasev_JPCM_2004}, and \cite{Oshima_APL_2003},
respectively.}

\begin{ruledtabular}
\begin{tabular}{l c c c c c}
                        &                       &  PBE   &  PBE0   &  Mixed    &  Expt.  \\
\hline
Si/SiO$_2$              & $\Delta E_{\text{v}}$ &  2.5   &  3.3    &  4.4      &  4.4 \\
                        & $\Delta E_{\text{c}}$ &  2.3   &  2.7    &  3.4      &  3.4 \\ 
4\emph{H}-SiC/SiO$_2$   & $\Delta E_{\text{v}}$ &  1.4   &  2.0    &  3.0      &  2.9 \\
                        & $\Delta E_{\text{c}}$ &  1.7   &  2.0    &  2.6      &  2.7 \\
Si/HfO$_2$              & $\Delta E_{\text{v}}$ &  2.3   &  3.1    &  2.9      &  2.9 \\
                        & $\Delta E_{\text{c}}$ &  1.5   &  1.9    &  1.7      &  1.7 \\
\end{tabular}
\end{ruledtabular}
\label{tab2}
\end{table}

\begin{figure}
\includegraphics[width=8.5cm]{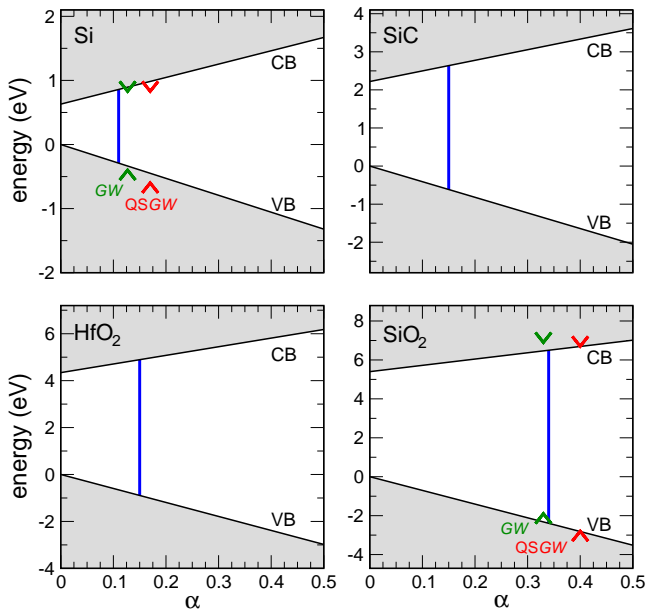}
\caption{(color online) Valence band maximum and conduction band minimum vs 
$\alpha$ for four different materials. Shifts calculated with $GW$ and 
quasiparticle self-consistent $GW$ (QS$GW$) \cite {Shaltaf_PRL_2008} are also shown.
Vertical lines represent the experimental band gaps and are shown 
in correspondence of $\alpha_0$ (see text).}
\label{fig2}
\end{figure}

To further improve band offsets, it appears imperative to more accurately describe 
the band gaps of bulk components.  This cannot be achieved for both interface components 
through the use of a hybrid functional with a fixed fraction of Hartree-Fock exchange 
$\alpha$. However, it has been argued that there is no universal fraction $\alpha$ 
for all materials and that its optimal value could even be property dependent 
\cite{Ernzerhof_IJQC_1997}. In Fig.\ \ref{fig2}, we show the evolution of the conduction 
and valence band edges of the four bulk materials considered in this work as a function of $\alpha$.
Since the bulk band gaps increase linearly with $\alpha$, the value of $\alpha$ can 
always be chosen to reproduce the experimental band gap. This resulted in optimal values 
$\alpha_0$ of 0.11 for Si, 0.15 for SiC and HfO$_2$, and 0.34 for SiO$_2$ 
(Table \ref{tab1} and Fig.\ \ref{fig2}). We note that for these $\alpha_0$ values,
the hybrid functional calculation not only gives the experimental band gap but also 
the position of the band extrema with respect to the adopted reference potential.
  
The consideration of a fraction $\alpha$ of the Hartree-Fock exchange 
is equivalent to an effective static screening of the long-range interaction: 
$\alpha\sim 1/\epsilon_\infty$, where $\epsilon_\infty$ is the electronic part 
of the dielectric constant. Indeed, the optimal $\alpha_0$'s found above respect 
this relationship in a qualitative way.  To further support the proposed adjustment 
of $\alpha$, we compared the band shifts of Si and SiO$_2$ calculated with the hybrid 
functionals to those obtained with $GW$ and quasiparticle 
selfconsistent $GW$ (QS$GW$) \cite{Shaltaf_PRL_2008}. The $GW$ band shifts are 
reported in Fig.\ \ref{fig2} in correspondence of the value of $\alpha$ which 
reproduces the band gap found in the $GW$ calculation. The comparison between 
hybrid and $GW$ band shifts is quite good, with differences not exceeding 
those between different $GW$ schemes.  


\begin{figure}
\includegraphics[width=7.5cm]{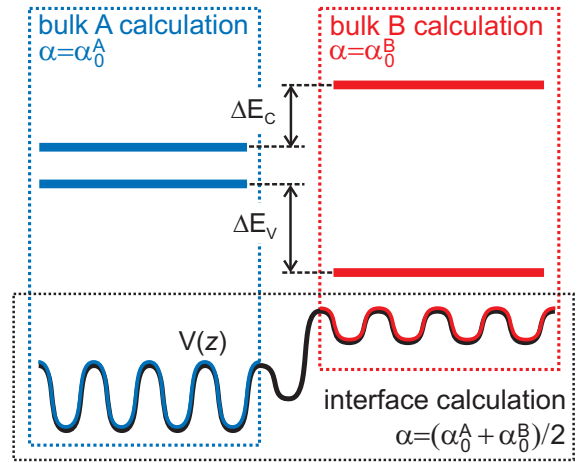}
\caption{(Color online) Schematic illustration of band offset determination at the interface 
between components $A$ and $B$ through the mixed scheme (see text). The local 
potential $V(z)$ is obtained from an interface calculation based on a hybrid functional with 
$\alpha=(\alpha_0^A+\alpha_0^B)/2$ and determines the lineup of the reference levels 
in the two components. For each component, the band extrema are then aligned
to the reference potential through bulk calculations based on  hybrid functionals
with a material-specific $\alpha_0$ chosen to reproduce its experimental band gap.
$\Delta E_{\rm v}$ and $\Delta E_{\rm{ c}}$ are the resulting valence and conduction 
band offsets.}
\label{fig3}
\end{figure}

The observations above can be combined to obtain band offsets. 
From the interface model, the lineup of the reference potentials in the 
two bulk components is determined. Then, bulk extrema are positioned 
on each side of the interface using the results of hybrid functional 
calculations with optimal $\alpha_0$ for each bulk component \cite{note3}. 
This mixed scheme is graphically illustrated in Fig.\ \ref{fig3}.

\begin{figure}
\includegraphics[width=6.0cm]{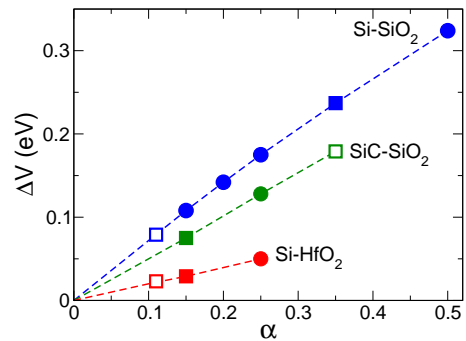}
\caption{(Color online) 
Lineup of the local potential (filled circles), expressed as a difference with respect 
to the result obtained with PBE ($\alpha$=0), vs fraction of Hartree-Fock 
for the Si-SiO$_2$, SiC-SiO$_2$, and Si-HfO$_2$ interfaces.
Squares are shown in correspondence of the $\alpha_0$'s of the individual 
bulk components (0.11 for Si, 0.15 for SiC and HfO$_2$, and 0.34 for SiO$_2$).
Filled symbols correspond to actual calculations, while open symbols are positioned
by interpolation or extrapolation.
}
\label{fig4}
\end{figure}

The scheme critically relies on the fact that the lineup extracted from the
interface calculation is not significantly dependent on the adopted fraction of 
Hartree-Fock exchange $\alpha$. To further support this point, we performed
several hybrid functional calculations with varying $\alpha$ for each of
the model interfaces under consideration. As shown in Fig.\ \ref{fig4},
the lineup between the reference levels in the two bulk components 
only marginally depends on $\alpha$, generalizing the observation made above
for the Si-SiO$_2$ interface. The dependence on $\alpha$ is even weaker 
for the other interfaces. Inspection of Fig.\ \ref{fig1}(b) shows that the
charge transfer occurring at the interface is similar to that observed in the 
oxide, suggesting that the change of the interfacial dipole should be attributed 
to a modification of the local chemistry rather than to the increase of the number 
semiconductor-induced gap states \cite{Tersoff_PRB_1984} resulting from the modified 
band offsets.  Assuming that the interface calculation is 
performed with a value of $\alpha$ corresponding to the average of the 
$\alpha_0$'s pertaining to the two interface components, we estimate that 
the induced indetermination is smaller than 0.15 eV in the worst case.

The band offsets obtained through the application of the mixed scheme are included 
in Table \ref{tab2}. The calculated values agree extremely well with experimental ones.
The error between theory and experiment is similar to the intrinsic indetermination
of our approach and to the scatter between different experiments. In comparison
to hybrid calculations with a fixed fraction $\alpha$, the mixed scheme provides 
a striking improvement in the theoretical estimation of band offsets. 


In conclusion, we demonstrated the accurate determination of band offsets through the
use of hybrid density functionals and experimental band gaps. The presented scheme 
constitutes a predictive tool which is computationally less demanding than $GW$ calculations, 
yet achieves band offsets of comparable accuracy \cite{Shaltaf_PRL_2008}.
Its application to complex interface components such as noncrystalline oxides is within 
reach without loss of accuracy.

Support from the Swiss National Science Foundation (Grants Nos.\
200020-111747 and 200020-119733) is acknowledged. The calculations
were performed on the BlueGene of EPFL, at DIT-EPFL and CSCS.


\begin{thebibliography}{10}
\bibitem{note3}
When a Kohn-Sham level is used as reference potential, the bulk 
calculation is first performed with the same $\alpha$ as in the 
interface calculation. The band edges are then adjusted 
according to Fig.\ \ref{fig2}.
\end{thebibliography}
\end{document}